\documentstyle[pre,aps,floats,twocolumn]{revtex}  
\input psfig  

\begin{document}

\title{Error propagation in the hypercycle}

\author{P. R. A. Campos and J. F. Fontanari}
\address{Instituto de F\'{\i}sica de S\~ao Carlos,
 Universidade de S\~ao Paulo\\
 Caixa Postal 369, 
 13560-970 S\~ao Carlos SP, Brazil 
}

\author{P. F. Stadler}
\address{Institut f\"ur Theorestische Chemie, 
 Universit\"at Wien\\ 
 W{\"a}hringerstra{\ss}e 17, A-1090 Wien,  Austria\\
 The Santa Fe Institute, 1399 Hyde Park Rd., Santa Fe NM 87501, USA
}


\maketitle

\begin{abstract}
We study analytically the steady-state regime of a network of $n$
error-prone self-replicating templates forming an asymmetric hypercycle and
its error tail.  We show that the existence of a master template with a
higher non-catalyzed self-replicative productivity, $a$, than the error
tail ensures the stability of chains in which $m < n -1$ templates coexist
with the master species.  The stability of these chains against the error
tail is guaranteed for catalytic coupling strengths ($K$) of order of $a$.
We find that the hypercycle becomes more stable than the chains only for
$K$ of order of $a^2$. Furthermore, we show that the minimal replication
accuracy per template needed to maintain the hypercycle, the so-called
error threshold, vanishes like $\sqrt{n/K}$ for large $K$ and $n \leq 4$.
\end{abstract}

\begin{abstract}
\end{abstract}

\section{Introduction}\label{sec:level1}

The limitation of the length of a genome by the replication accuracy
per nucleotide ($q$) has led to a deadlock in the theories of the
origin of life based on the evolution of competing self-replicating
polynucleotides. 
According to Eigen's quasispecies model \cite{eigen,review}, which may
serve as a paradigm here, polynucleotides have to replicate with high
accuracy in order to reach a certain length, a requirement that is
impossible to fulfill without the aid of specialized catalysts. However, to
build those catalysts a blueprint is necessary that amounts to a large
genome (the nucleotide sequence), which itself cannot be maintained without
the catalysts. In particular, for polynucleotides of fixed length $L$, the
quasispecies model predicts the existence of a minimal replication accuracy
per genome $Q_c =q_c^L$, below which the genetic information is
irreversibly lost. This information crisis has been termed error threshold
transition. Above $Q_c$ the population is composed of a master copy
together with a cloud of structurally similar mutants (quasispecies)
\cite{eigen,review}. Equally important is the finding that, except in a
trivially degenerate case, two or more quasispecies cannot coexist
\cite{swetina}, thus precluding the coexistence of templates
(i.e. polynucleotides) sufficiently different from each other to code for
any useful set of catalysts. Although it has been claimed that the
information crisis is not really a fundamental issue, since the error
threshold transition appears only in some pathological, discontinuous
replication landscapes \cite{wiehe}, the coexistence problem seems to be
more pervasive, as it is associated to the form of the growth functions in
the chemical kinetics equations \cite{tree,wills}.

In order to circumvent the aforementioned limitations of the quasispecies
model, Eigen and Schuster proposed the hypercycle \cite{hyper}, that is, a
catalytic feedback network whereby each template helps in the replication
of the next one, in a regulatory cycle closing on itself. This model has
gain plausibility when the ability of polynucleotides to help propagate
each other was established experimentally through the study of the
catalytic activity of the RNA (ribozymes) \cite{doudna,bio}. Interestingly,
though the error threshold phenomenon has traditionally been considered the
main motivation for the proposal of the hypercycle (see \cite{maynard}, for
instance), most of the seminal works in this field have dealt with the
coexistence issue only, as they assume perfect replication accuracy for the
hypercycle elements \cite{hyper,hofbauer}. In this case an arbitrary number
of templates permanently coexist in a dynamical equilibrium state; if $n>4$,
however, the template concentrations vary with time \cite{hyper},
periodically  decreasing to very small values. In practice,
large hypercycles are therefore susceptible to extinction via fluctuations,
see e.g.\ \cite{nuno0}, hence the information gain due to the coexistence
of different templates in the hypercycle may not be very impressive after
all. Furthermore, we will argue in this paper that coexistence in the
absence of a stable equilibrium can also be achieved by a simpler
arrangement, namely, the free chains, in which the cyclic order of the
catalysts is interrupted.

The effect of error-prone replication (mutation) in the hypercyclic
organization was investigated by introducing a mutation field as a
perturbation of the error-free kinetic equations \cite{stadler}. This
approach, however, is not very appropriate to study the error threshold
phenomenon, since the results obtained cannot be easily compared with those
of the quasispecies model.  In this sense, a better approach is to assume
the existence of a special class of templates with no catalytic activity,
so-called error-tail, that appear as a consequence of the replication
errors of the hypercycle elements \cite{garcia,nuno1,nuno2}. However, the
particular catalytic network investigated extensively within that framework
was not the hypercycle, except for a short discussion in \cite{nuno3}, but
the fully connected network in which each element helps the replication of
all other elements of the network \cite{nuno1,nuno2}.  (Clearly, in the
case of $n=2$ elements, these two networks become identical \cite{garcia}.)
Such a network is more robust than the hypercycle since the mal-functioning
or extinction of one of its elements does not compromise the whole network.
Nevertheless, besides its aesthetic appeal, the cyclic coupling of the
hypercycle seems to be more realistic \cite{bio}.

The goal of this paper is to investigate analytically the
steady-states of a deterministic system comprised of two parts,
namely, a hypercycle made up of $n$ self-replicating templates $I_1,
I_2, \ldots, I_n$ and its error tail $I_e$. These parts are coupled
such that any erroneous copy of the hypercycle elements will belong to
the error tail.  The focus of the present analysis is on the location
in the parameters space of the model (i.e. replication accuracy per
template, non-catalyzed and catalyzed productivity values, and
hypercycle size) of the regions of stability of the diverse
possibilities of coexistence between the templates composing the
hypercycle. In particular, we give emphasis to the characterization of
the critical parameters at which the hypercycle becomes unstable
against the error tail.

The remainder of the paper is organized as follows. In Sec.\
\ref{sec:level2} we present the chemical kinetics equations that
govern the time evolution of the system and motivate the specific
choice of the parameters used throughout the paper. The fixed-points
of the kinetic equations are obtained analytically in Sec.\
\ref{sec:level3} and their stability discussed in Sec.\
\ref{sec:level4}. The phase-diagrams showing the regions of stability
of the diverse coexistence states are presented and analyzed in Sec.\
\ref{sec:level5}.  Finally, some concluding remarks are presented in
Sec.\ \ref{sec:level6}.

%
\section{The model}\label{sec:level2}
%

\begin{figure}
\psfig{file=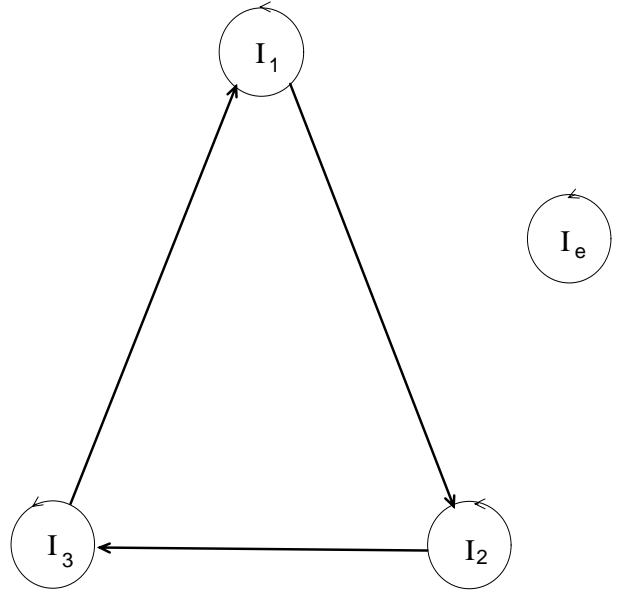,width=0.47\textwidth}
\caption{The system composed of a hypercycle of size $n=3$
and its error tail $I_e$. The thin arrows represent the 
non-catalyzed self-replication reactions and the thick arrows 
represent the self-replication catalytically assisted by
the neighbor template}
\label{Fig.1}
\end{figure}

We consider a system composed of a hypercycle made up of $n$ elements $I_1,
\ldots, I_n$ and its error tail $I_e$, as illustrated in
Fig.~\ref{Fig.1}. In contrast to the so-called elementary hypercycle
\cite{hyper}, we assume that the templates are capable of self-replication
with productivity values $A_i ~\left( i=1, \ldots, n \right )$ and
$A_e$. Moreover, as usual, the growth promotion of template $I_i$ as a
result of the catalysis from template $I_{i -1}$ is measured by the kinetic
constants $K_i$. The key ingredient in the modeling is that in both
processes of growth of template $I_i$ the probability of success is given
by the parameter $Q \in [0,1]$, so that an erroneous copy, which will then
belong to the error tail, is produced with probability $1-Q$.  Hence the
concentrations $x_i~\left ( i=1, \ldots,n \right )$ of the hypercycle
elements and the concentration $x_e$ of the error-tail evolve in time
according to the kinetic equations
\begin{equation}\label{ode_h}
\dot x_{i} = x_{i} \left ( A_{i}Q + K_i x_{i-1}Q - \Phi \right ) 
~~~~~~~ i=1,...,n
\end{equation}
and
\begin{equation}\label{ode_e}
\dot x_{e} = x_{e} \left ( A_{e} - \Phi \right ) + \left ( 1-Q \right ) 
\sum_{i=1}^{n} x_{i} \left ( A_{i}  + K_i x_{i-1} \right )  
\end{equation}
where $x_0 \equiv x_n$ and
\begin{equation}\label{flux}
\Phi= \sum_{i=1}^n x_i \left ( A_i + K_i x_{i-1} \right ) + A_e x_e  
\end{equation}
is a dilution flux that keeps the total concentration constant, i.e.,
$ \sum_{i=1}^{n} \dot x_{i} + \dot x_{e}  = 0$. As usual,
the dot denotes a time derivative. Henceforth we will assume
that \cite{nuno1,nuno2}
\begin{equation}\label{CC}
 \sum_{i=1}^{n} x_{i} +x_{e} = 1 .
\end{equation}
Clearly, this formulation is equivalent to considering polynucleotides
of length $L \rightarrow \infty$ whose replication accuracy per
nucleotide $q$ goes to 1 such that the replication accuracy per genome
is finite, i.e.  $q^L \rightarrow Q$. In this limit, the
back-mutations from the error-tail elements to the templates that
compose the hypercycle, as well as the mutations between those
templates, can be safely neglected.  Hence, mutations can only
increase the concentration of the elements in the error-tail.  The
advantage of working in this limit is that the error threshold
transition can be precisely located by determining the value of $Q$ at
which the concentration of a relevant template vanishes. For finite
$L$, as well as for finite population sizes, the characterization of
this transition is more involved, being achieved through the use of
finite-size scaling techniques \cite{prac}.

In this work we consider the single-sharp-peak replication landscape
\cite{eigen,review}, in which we ascribe the productivity value
$A_{1}=a>1$ to the so-called master template $I_1$ and $A_{i} = A_e=1$
to the $n-1$ other elements of the hypercycle as well as to the
error-tail. Also, for the sake of simplicity we set $K_i = K$ for all
$i$. The motivation for this particular choice of parameters is the
observation that the emergence of the hypercycle requires both spatial
and temporal coexistence of the templates forming the network, and
this can be achieved by a quasispecies distribution, which guarantees
the coexistence of the master template and its close mutants, despite
the purely competitive character of the quasispecies model
\cite{emergence}.  Once the coexistence is established, the appearance
of catalytic couplings between the templates is not a very unlike
event. Of course, as soon as those cooperative couplings become
sufficiently strong to balance the competition imposed by the constant
concentration constraint, the mutants will certainly depart from the
master template due to the relentless pressure of the mutations, so
that no trace will remain of the original quasispecies distribution.

%
\section{Fixed points}\label{sec:level3}
%

Let us distinguish between surviving templates $x_i>0$ and extinct
templates $x_i=0$.  A survivor $I_j$ is said isolated if
$x_{j-1}=x_{j+1}=0$. Hence,
\begin{equation}\label{j}
\dot x_j = x_j \left ( Q - \Phi \right )  ~~~~ j > 1
\end{equation}
and
\begin{equation}\label{e}
\dot x_e = x_e(1\!-\!\Phi)\!+\!(1\!-\!Q)\!  
\left[x_1\left(a\!-\!1\right)\!+\!1\!+\! 
K\!\!\sum_{i\neq j,j+1}\!\!\! x_{i}x_{i-1}\right].
\end{equation}
In the steady-state regime, Eq.\ (\ref{j}) yields $\Phi = Q$ which,
for $Q < 1$, is incompatible with Eq.\ (\ref{e}), since the term
within brackets in this equation is positive. Therefore, all isolated
survivors with the exception of the master template are unstable
against the error tail.  Next consider the following chain of
surviving templates:
\begin{equation}
\dot x_i     = x_i( Q - \Phi ) 
\end{equation}
\begin{equation}
\dot x_{i+1} = x_{i+1}( Q +KQx_{i} - \Phi ) 
\end{equation}
\begin{equation}
\dot x_{i+2} = x_{i+2}( Q +KQx_{i+1} - \Phi ) 
\end{equation}
\begin{equation}
\vdots
\end{equation}
\begin{equation}
\dot x_{k}   = x_{k}( Q +KQx_{k-1} - \Phi ) 
\end{equation}
which does not contain $x_1$. Again, in the steady-state regime the
first equation yields $\Phi = Q$, implying $KQx_{i}=0$, i.e.,
$x_{i}=0$. So, there is no fixed point corresponding to such a
chain. Any chain of survivors therefore must start with template $n$
or $1$. In the first case we get $ \Phi = Q$ from $\dot x_n=0$ and
$aQ+KQx_n=Q$ from $\dot x_1=0$, yielding $x_n=(1-a)/K<0$, which rules
out this possibility.  The equilibria of interest for our study thus
are either the interior equilibrium in which all templates survive, or
a fixed point that corresponds to a chain of survivors beginning with
$I_1$.

Accordingly, we define a $m$-coexistence state as the $n$-component
template vector ${\bf x} = \left( x_1, x_2, \ldots, x_n \right )$ in
which the first $m$ components are strictly positive and the rest are
equal to zero. Clearly, given the template vector ${\bf x}$ the
concentration of the error tail $x_e$ is determined by the constraint
(\ref{CC}).  In the following we solve analytically the kinetic
equations in the steady-state regime $\dot x_i = 0 ~\forall i$.

The simplest fixed point is the zero-coexistence state ($m=0$) which
corresponds to the solution $x_1=\ldots=x_n = 0$ and $x_e= 1$,
existing for the complete range of parameter values.

In the case of chains, i.e. $0 < m < n$, the steady-state solutions of
Eqns. (\ref{ode_h}) and (\ref{ode_e}) are straightforward. In fact,
since $x_n = 0$ by definition, we get $\Phi = aQ$ from $\dot x_1 = 0$
which then yields
\begin{equation}
x_1 = x_2 = \ldots = x_{m-1} = \frac{a -1}{K}.
\end{equation}
Next we insert this result in Eq.\ (\ref{flux}) to obtain
\begin{equation}\label{xm}
x_m = \frac{Qa-1}{a-1} - \left( m-1 \right) \frac{a-1}{K}.
\end{equation}
However, since $x_i \in \left ( 0,1 \right )~\forall i$, this solution
is physically meaningful in the region $K > a-1$ and $Q > Q_m$ where
\begin{equation}\label{Qm}
Q_m = \frac{1}{a} + \frac{1}{Ka}~ \left ( m-1 \right ) \left ( a-1 \right )^2 .
\end{equation}
We note that the $1$-coexistence state (quasispecies) is obtained by
setting $m=1$ in Eq.\ (\ref{xm}) and its region of existence is simply
$Q > 1/a$, since the other condition, namely $K > a-1$, is derived by
considering the other templates in the chain. In fact, this very
simple result quantifies nicely the notion that the cooperative
couplings must reach a certain minimum strength so as to balance the
competition between templates.

The analysis of the hypercycle, i.e. $m=n$, is a little more involved.
>From $\dot x_2 = 0$ we get $\Phi - Q = K Q x_1$ which, inserted in the
equations $\dot x_3 = \ldots = \dot x_n=0$, yields $x_1 = x_2 = \ldots
= x_{n-1}$ and
\begin{equation}\label{co_n}
x_n = x_1 - \frac{a-1}{K} .
\end{equation}
Finally, using these results in Eq. (\ref{flux}) we find that $x_1$
is given by the roots of the quadratic equation
\begin{equation}\label{quad}
n K x_1^2 - \left ( K Q + a -1 \right ) x_1 + 1 - Q = 0 .
\end{equation}
For $ K < \left ( a -1 \right )^2/4n$, this equation has real roots
for all $Q \geq 0$, otherwise it has real roots for $Q \geq Q_h$ where
$Q_h$ is the unique positive root of the equation
\begin{equation}\label{Qh}
K^2 Q_h^2  + 2 K  \left(  a -1 + 2 n \right ) Q_h +
\left ( a -1 \right )^2 - 4 n K = 0 .
\end{equation}
In particular, for large $K$ we find $Q_h \approx 2 \sqrt{n/K}$.
Furthermore, it can be easily seen from Eqns.\ (\ref{co_n}) and
(\ref{quad}) that $x_n$ vanishes at $Q = Q_n$ with $Q_n$ given in Eq.\
(\ref{Qm}). To understand the role of $Q_h$ and $Q_n$ (we note that
$Q_n \geq Q_h$) in delimiting the region of existence of the
$n$-coexistence state we must look at the behavior of the two real
roots of Eq.\ (\ref{quad}).  Let us denote them by $x_1^+$ and $x_1^-$
with $x_1^+ \geq x_1^-$, which, according to Eq.\ (\ref{co_n}),
correspond to $x_n^+$ and $x_n^-$, respectively.  Of course, these
roots become identical at $Q=Q_h$ and so the two solutions for $x_n$
will vanish simultaneously only at the value of $K=K_h$ at which $Q_h$
equals $Q_n$. Explicitly, we obtain
\begin{equation}\label{Kh}
K_h = \left ( a-1 \right ) \left [ n \left ( a+1 \right ) - 1 \right ] .
\end{equation}
by inserting Eq.\ (\ref{Qm}) into Eq.\ (\ref{Qh}).  Although both roots
$x_1^+$ and $x_1^-$ are in the simplex $\left ( 0,1 \right )$, this is not
so for $x_n^+$ and $x_n^-$. In particular, for $K < K_h$ both
concentrations are negative within the range $Q_h \leq Q < Q_n$.  However,
while $x_n^+$ becomes positive for $Q > Q_n$ (it vanishes at $Q_n$),
$x_n^-$ remains always negative.  Since $K_h > \left ( a -1 \right )^2/4n$
the same conclusion holds in the range $K < \left ( a -1 \right )^2/4n$ as
well, provided we define $Q_h = 0$ in this region.  The situation is
reversed for $K > K_h$: both concentrations are positive within the range
$Q_h \leq Q < Q_n$, but now it is $x_n^-$ that vanishes at $Q_n$ and
becomes negative for $Q > Q_n$ while $x_n^+$ remains always positive.
Despite the small region in the parameters space where the root $x_1^-$
yields concentrations inside the simplex, the linear stability analysis
discussed in the sequel indicates that this solution is always unstable, so
we only need to consider the root $x_1^+$.  Thus the range of existence of
the hypercycle fixed point $m=n$ is $Q \geq Q_n$ if $K \leq K_h$ and $Q
\geq Q_h$ if $K > K_h$.

In models without error-tail, i.e., pure replicator equations a much
stronger statement on coexistence is possible. The ``time average theorem''
\cite{hs-book} states that if there is a trajectory along which a certain
subset of templates $J$ survives, then there is a fixed point with exactly
the $J$-coordinates non-zero. While we have not been able to prove the
``time average theorem'' in full generality for Eqns.~(\ref{ode_h}) and
(\ref{ode_e}), it is easily verified for free chains. Hence, if there is
no $m$-coexistence equilibrium, then there is no trajectory at all along
which the templates $I_1$ through $I_m$ survive. 

Hitherto we have determined the ranges of the parameter $Q$ where the
$m$-coexistence states are physically meaningful, in the sense that
$x_i \in \left ( 0,1 \right )~\forall i$. The next step is to find the
regions where these states are locally stable.

%
\section{Stability analysis}\label{sec:level4}
%

In order to perform a standard linear stability analysis of the fixed
points obtained in the previous section, it is convenient to rewrite
the kinetic equations (\ref{ode_h}) and (\ref{ode_e}) as follows
\begin{equation}
\dot x_i = x_i F_i \left ( {\bf x} \right ) ~~~~i=1, \ldots,n
\end{equation}
where
\begin{equation}\label{F}
F_i({\bf x}) =  A_i Q + K Q x_{i-1} - A_e -\!\sum_j x_j
\left(A_j\!-\!A_e\!+\!K x_{j-1}\right) 
\end{equation}
and we have used the constraint (\ref{CC}) to eliminate $x_e$.  The
stability of a fixed point is ensured provided that the real parts of
all eigenvalues of the $n \times n$ Jacobian ${\cal J}$ are negative.
In our case the elements of the Jacobian are given by
\begin{equation}\label{M}
J_{ij} = \delta_{ij} F_i + x_i \frac{\partial{F_i}}{\partial{x_j}}  
~~~~i,j=1,\ldots,n .
\end{equation}
The evaluation of the eigenvalues is simple only for the
zero-coexistence state, since in this case the Jacobian is diagonal
with elements $J_{11} = a Q - 1$ and $J_{ii} = Q-1,~ i > 1$. Therefore
this steady state becomes unstable for $Q > 1/a$, which coincides with
the lowest replication accuracy required for the existence of the
$1$-coexistence state. However, for a general $m$-coexistence state we
have to resort to a numerical evaluation of the Jacobian eigenvalues.

Fortunately, in the case of chains $0 < m < n$ there is an alternative
way to look at the stability of the fixed points, as hinted by the
stability analysis of the zero-coexistence state, which becomes
unstable due to the emergence of the $1$-coexistence state. In fact,
it can be easily seen that any perturbation of the $m$-coexistence
fixed point which makes the concentration $x_{m+1}$ non-zero will be
amplified if $A_{m+1}Q + K Q x_m - \Phi$ is positive. For $m>0$ we use
$\Phi = aQ$ and $A_{m+1} = 1$ together with the value of $x_m$ given
in Eq.\ (\ref{xm}) to obtain the following (necessary) condition for
the stability of the $m$-coexistence state,
\begin{equation}\label{stab_c}
Q < Q_{m+1}  ~~~~ m > 0,
\end{equation}
with $Q_m$ given in Eq.\ (\ref{Qm}).  Hence the maximum value of $Q$
allowed for the stability of the $m$-coexistence state coincides with
the minimum $Q$ required for the existence of the $(m+1)$-coexistence
state.  Interestingly, though for $m=0$ we have $\Phi = 1$, $A_1 = a$
and $x_0 = 0$, condition (\ref{stab_c}) holds true in this case too.

At this point two caveats are in order. First, the entire argument
leading to the stability condition (\ref{stab_c}) is flawed if the
$(m+1)$-coexistence state happens to be unstable. Therefore, we must
guarantee via the numerical evaluation of the Jacobian eigenvalues
that the $l$-coexistence state is stable before using that condition
to study the stability of chains with $m < l$.  In particular, we have
carried out the numerical analysis for the hypercycle solution $l=n$
and found the following general results:
\begin{itemize}
\item[(i)] For $ n \leq 4$, it  is always stable; 
\item[(ii)] for $ n =5$, it is stable in a very small range of $Q$ above 
$Q_5$; and 
\item[(iii)] for $n \geq 6$, it is always unstable.
\end{itemize}    
Second, the derivation of the stability condition (\ref{stab_c}) is
based on the analysis of a single eigenvalue of the Jacobian and so it
does not yield a sufficient condition for the stability of the fixed
points. Nevertheless, we have verified through the numerical
evaluation of all $n$ eigenvalues that, provided the
$(m+1)$-coexistence state is stable, the eigenvalue associated to
fluctuations leading to an increase of the chain length is the first
one to become positive.
 
%
\section{Discussion}\label{sec:level5}
%

Combining the existence and the stability results derived in the
previous sections we can draw the phase-diagrams in the plane $(K,Q)$
for fixed $a$ and $n$. In particular, for $n \leq 4$ the
$m$-coexistence state is stable within the interval
\begin{equation}\label{chain}
Q_m <Q<Q_{m+1} ~~~~~ m < n
\end{equation}
with $Q_m$ given in Eq.\ (\ref{Qm}), provided that $K > a-1$.
Interestingly, for fixed $Q$, Eq.\ (\ref{chain}) shows that the
increment $\delta K$ in the catalytic coupling needed to incorporate a
new template into the chain is
\begin{equation}\label{DK}
\delta K = \frac{ \left ( a - 1 \right)^2 }{aQ - 1} ,
\end{equation}
regardless the number of elements in the chain.  The case $ K < a -
1$, for which no chains with $m>1$ are allowed, does not require any
special consideration. In fact, we find that the only stable states
are the zero-coexistence ($Q < 1/a$) and the $1$-coexistence ($1/a
\leq Q \leq 1$) states. However, since $Q_2 \leq 1$ only for $K \geq a
-1$, this result is consistent with Eq.\ (\ref{chain}).

\begin{figure}
\psfig{file=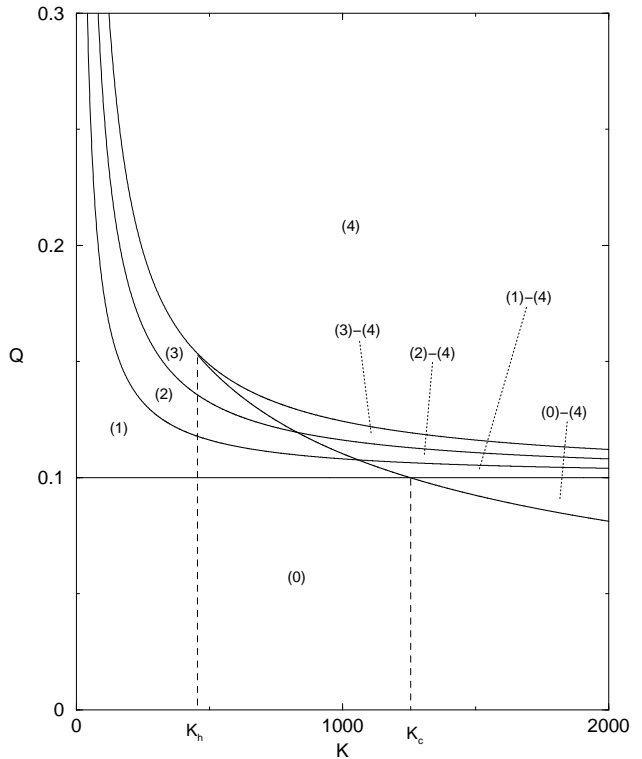,width=0.47\textwidth}
\caption{Phase-diagram in the space $\left ( K,Q \right )$ for $n=4$
and $a=10$ showing the regions of stability of the diverse coexistence
states.  The numbers between parentheses indicate the number of
coexisting templates.  Regions of bistability appear for $K >
K_h$. The thin lines are (from bottom to top) $Q_1, Q_2, Q_3$ and
$Q_4$. The thick line is $Q_h$.}
\label{Fig.2}
\end{figure}

The $n$-coexistence state (i.e., the hypercycle solution) is stable
for $Q > Q_n$ if $K \leq K_h$ and for $ Q > Q_h$, otherwise, where
$K_h$ and $Q_h$ are given by Eqns.\ (\ref{Kh}) and (\ref{Qh}),
respectively.  We define the error threshold of the hypercycle as the
value of the replication accuracy $Q$ that delimits the region of
stability of the $n$-coexistence state. The phase diagram for $a=10$
and $n=4$ shown in Fig.~\ref{Fig.2} illustrates the major role played by $K_h$
in the hypercyclic organization: only for $ K > K_h$ the hypercycle
becomes more stable than a chain of the same length.  Another
important quantity is the value of $K$, denoted by $K_c$, at which
$Q_h$ equals $1/a$, the minimal replication accuracy of the
quasispecies. It is given by
\begin{equation}\label{Kc}
K_c = a \left ( a - 1 \right ) \left [ 2n - 1 + \sqrt{n \left ( n - 1
\right)} \right ] .
\end{equation}
Beyond this value the error threshold of the hypercycle $Q_h$ is
smaller than that of the quasispecies.  Moreover, as mentioned before,
for large $K$, it vanishes like $1/\sqrt{K}$.  A rather frustrating
aspect of $K_h$ and $K_c$ is that both are of order $a^2$, indicating
then that the productivity of catalytically assisted self-replication
is much larger than that of non-catalyzed self-replication. While this
is obviously true for biochemical catalysis, it is difficult to argue
for the existence of such efficient catalysts in prebiotic conditions.
On the other hand, we can take a different, more optimistic viewpoint
and argue that modern biochemical catalysts (enzymes) are so efficient
because their precursors had to satisfy the stringent conditions
imposed by surpassing $K_h$.

\begin{figure}
\psfig{file=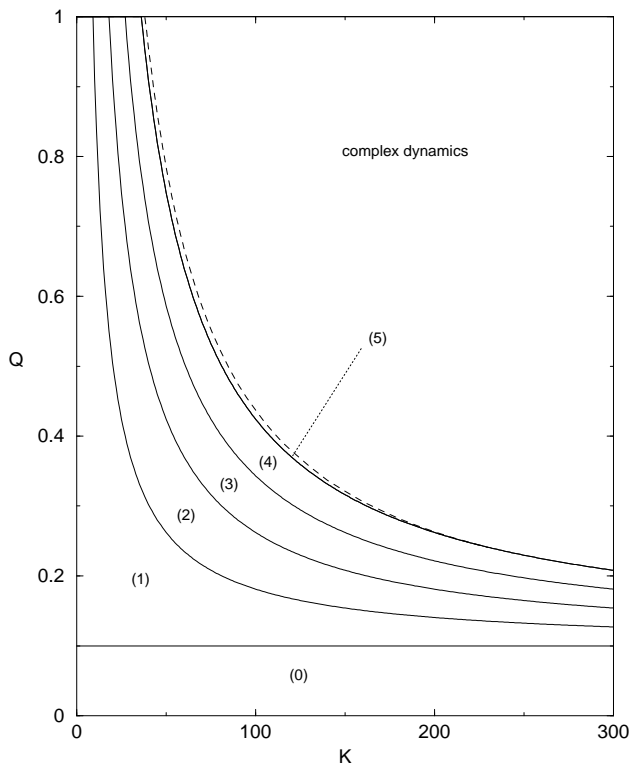,width=0.47\textwidth}
\caption{Same as Fig.~\ref{Fig.2} but for $n=5$. There are no stable fixed
points above the dashed curve. The solid curves are (from bottom to
top) $Q_1, Q_2, Q_3,Q_4,$ and $Q_5$.}
\label{Fig.3}
\end{figure}

\begin{figure}
\psfig{file=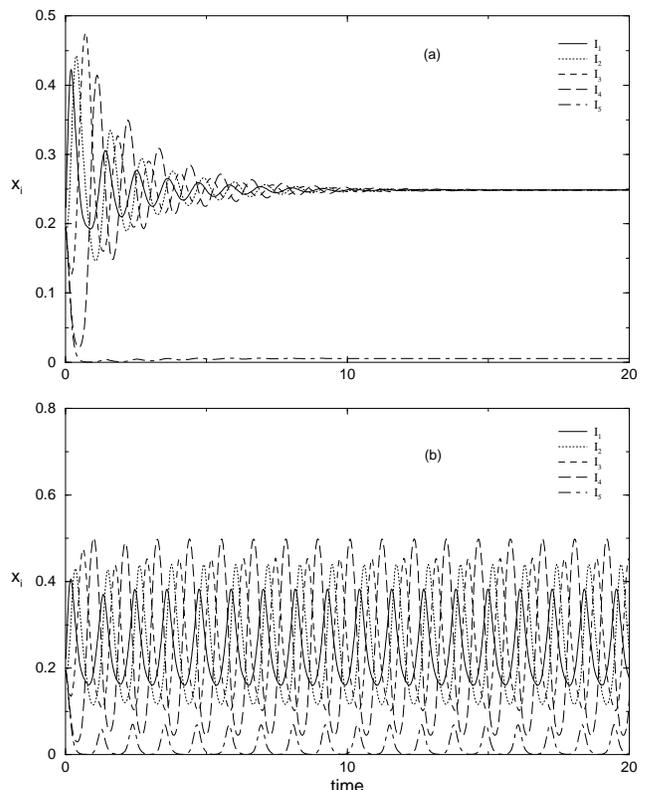,width=0.47\textwidth}
\caption{Time evolution of the five concentrations of the templates
composing a hypercycle of size $n=5$ for $a=10$, $Q= 1$, and (a) $K=
37$ (inside the region of stability) and (b) $K= 40$ (outside the
region of stability).  The initial state is $x_i (0) = 0.2 ~\forall
i$.}
\label{Fig.4}
\end{figure}

In Fig.~\ref{Fig.3} we present the phase-diagram for $n=5$. The main
difference from the previous figure is that the $5$-coexistence state is
stable only within the thin region between $Q_5$ and the dashed curve,
obtained through the numerical evaluation of the Jacobian eigenvalues. As
these curves intersect at some $K \leq K_h$, the $5$-membered hypercycle is
not very interesting, since it has the same characteristics of a chain of
length $m=5$.  To confirm this result we have carried out the numerical
integration of the kinetic equations using the ninetieth-order Runge-Kutta
method. The results are shown in Fig.~\ref{Fig.4}, which illustrates the
time evolution of the concentrations $x_i~ \left ( i=1, \ldots, 5 \right )$
inside and outside the region of stability. Although the behavior pattern
in the region of instability seems periodic, we have not explored
completely the space of parameters to discard the existence of chaotic
behavior. Hence we use the term complex dynamics to label this region in
Fig.~\ref{Fig.3}. We note that the phase-diagram shown in this figure
describes also the regions of stability of hypercycles and chains of size
$n \geq 5$, since $m$-coexistence states with $m >5$ are always unstable.

An interesting limiting case which deserves special attention is the
symmetric hypercycle ($a=1$).  According to the argument put forward
in the beginning of Sec.\ \ref{sec:level3}, the only fixed points in
this case are the zero-coexistence state and the hypercycle, i.e.,
chains are not allowed. Moreover, Eq.\ (\ref{co_n}) yields $x_1 = x_2
= \ldots = x_n$ where $x_1$ is given by Eq.\ (\ref{quad}) with $a$
replaced by $1$.  The analysis of the roots of that quadratic equation
and the numerical evaluation of the Jacobian eigenvalues yield that
the symmetric hypercycle is stable for
\begin{equation}
K > \frac{4 n}{Q^2}~\left ( 1 - Q \right ) , 
\end{equation}
provided that $n \leq 4$. The region of stability observed in Fig.~\ref{Fig.3}
for the $5$-coexistence state does not appear in the symmetric case
$a=1$, so it must be a consequence of the asymmetry in the
productivity values of the non-catalyzed self-replication reaction. We
note that, differently from the asymmetric case ($a > 1$), the
zero-coexistence state is always stable.

For the sake of completeness, we present some results on the
elementary hypercycle ($A_i = 0, ~i=1, \ldots,n$) coupled to an error
tail ($A_e = 1$) via the imperfect catalytically assisted
self-replication. Inserting these parameters into Eq.\ (\ref{F}) and
setting $F_i = 0~\forall i$ yields $x_1 = \ldots = x_n$ with $x_1$
given by the larger root of the quadratic equation
\begin{equation}
K n x_1^2 - \left ( n + K Q \right ) x_1 + 1 = 0 ,
\end{equation}
since we have verified that the smaller root is always unstable.  As
in the symmetric case discussed above, for $ n \leq 4$ the stability
condition coincides with the condition for real $x_1$, namely,
\begin{equation}\label{ele}
Q \geq 2 \sqrt{\frac{n}{K}} - \frac{n}{K}.
\end{equation}
Thus the term in the right hand side of this inequality yields the
error threshold of the elementary hypercycle.

\begin{figure}
\psfig{file=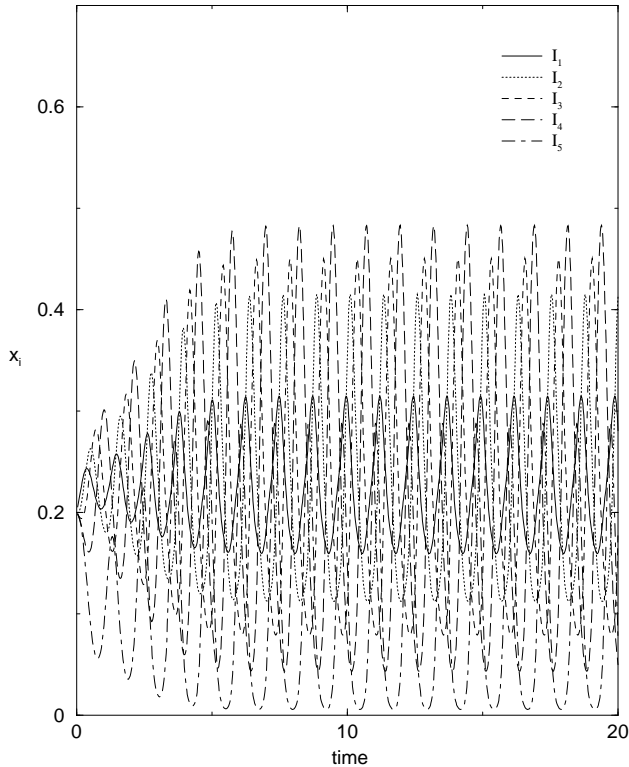,width=0.47\textwidth}
\caption{Time evolution of the five  concentrations of the templates
composing a free chain of size $n=5$. The parameters and initial state
are the same as for Fig.~\ref{Fig.4}(b).} 
\label{Fig.5}
\end{figure}

Before concluding this section, we must note that the chains of size $m$
considered hitherto are bonded in a hypercycle of size $n > m$.  We could
study {\it free} chains of size $n$ as well by simply setting $x_0 = 0$ in
the kinetic equations (\ref{ode_h}) and (\ref{ode_e}).  Not surprisingly,
the results are essentially the same as for bonded chains, with $Q_m$
playing a similar fundamental role in delimiting the regions of stability
of the shorter chains ($m < n$). Although a full discussion of the
stability of the complete chain ($m=n$) is beyond the scope of this paper,
we point out that free chains with $n>4$ are always unstable. Moreover, as
illustrated in Fig.~\ref{Fig.5}, the oscillatory behavior pattern of the
template concentrations, which ensues a dynamic coexistence among all
templates in the chain, is similar to that observed in the hypercycle
(compare with Fig.~\ref{Fig.4}).  In this sense, the free chains seem as
good as the hypercycle to attain that kind of coexistence. However, we must
emphasize that, for sufficiently large $K$ (i.e. $K > K_h$), the fixed
points describing the coexistence of $n \leq 4$ templates in a hypercycle
are much more robust against replication errors than their counterparts in
the free chains.

\section{Conclusion}\label{sec:level6}
 
Our study of the steady states of an asymmetric hypercycle composed of
$n$ error-prone self-replicating templates indicates that, for $n \leq
4$, the error threshold of the hypercycle ($Q_h$) becomes smaller than
that of the quasispecies ($Q_1 = 1/a$) for catalytic couplings ($K$)
of order of $a^2$, where $a$ is the productivity value of the master
template. In particular, $Q_h$ vanishes like $\sqrt{n/K}$ for large
$K$. Perhaps, even more important is our finding that the asymmetry in
the non-catalyzed self-replication reaction ($a > 1$) entails the
existence of chains of size $n \leq 5$. We note that these chains are
unstable in the symmetric hypercycle as well as in the fully connected
network \cite{nuno1,nuno2}.
 
Adding to the scenario for the emergence of the hypercycle described
in Sec.\ \ref{sec:level2}, which starts with an isolated quasispecies,
our results indicate that the chains may well be the next step in this
complex evolutionary process.  In fact, according to Eq.\ (\ref{DK})
the strengths of the catalytic couplings needed to form a chain are of
order $a$, while the hypercycle only acquires its desirable stability
characteristics for strengths of order $a^2$ (see Eq.\ (\ref{Kh})).
Although we realize that an evolutionary step leading from chains to
hypercycles is still a major one, it is certainly much more plausible
than a direct transition from quasispecies to hypercycle. In any
event, we think that the emergence of the hypercycle can be explained
as a series of plausible smooth transitions, without need to
postulating the hypercycle as an unique event in prebiotic
evolution. In this vein, this work represents a modest first step to
tackle this fundamental problem within a firm basis.
 
To conclude we must mention an alternative resolution for the
information crisis in prebiotic evolution which has received some
attention recently \cite{book}, namely, the stochastic corrector model
\cite{Szat}. This model builds on ideas of the classical group
selection theory for the evolution of altruism \cite{altru}, since it
considers replicative templates competing inside replicative
compartments, whose selective values depend on their template
composition.  However, the chemical kinetics equations governing the
dynamics of the templates inside the compartments display a
non-physical (non-integer exponents) dependence of growth on template
concentrations. It seems to us that this basic assumption of the
stochastic corrector model must be relaxed, or at least justified,
before it can be considered an important alternative to the more
traditional approach based on the hypercycle and its variants.

\bigskip

P.R.A.C.\ thanks Prof.\ P.\ Schuster and Prof.\ P.\ F.\ Stadler for
their kind hospitality at the Institut f\"ur Theorestische Chemie, where
part of his work was done, and Prof.\ P.\ E.\ Phillipson for illuminating
discussions.  The work of J.\ F.\ F.\ was supported in part by Conselho
Nacional de Desenvolvimento Cient{\'\i}fico e Tecnol{\'o}gico (CNPq).
The work of P.F.S.\ was supported in part by the Austrian 
Fonds zur F{\"o}rderung der Wissenschaftlichen Forschung, Proj.\ No.\
13093-GEN. P.R.A.C.\ is supported by FAPESP.

\end{document}